\documentclass[11pt]{article}
\usepackage{graphicx}
\usepackage{amssymb}
\usepackage{graphics}

\input{tcilatex}

\begin{document}

\title{Comment about constraints on nanometer-range modifications to gravity
from low-energy neutron experiments}
\author{O. Zimmer and N. Kaiser \\
%EndAName
Physik-Departments E18 and T39, Technische Universit\"{a}t M\"{u}nchen, \\
85748 Garching, Germany}
\maketitle

\begin{abstract}
A topic of present interest is the application of experimentally observed
quantum mechanical levels of ultra-cold neutrons in the earth's
gravitational field for searching short-range modifications to gravity. A
constraint on new forces in the nanometer-range published by Nesvizhevsky
and Protasov follows from inadequate modelling of the interaction potential
of a neutron with a mirror wall. Limits by many orders of magnitude better
were already derived long ago from the consistency of experiments on the
neutron-electron interaction.\medskip


email: oliver.zimmer@ph.tum.de, norbert.kaiser@ph.tum.de
\end{abstract}

In a recent experiment, quantum mechanical levels of the neutron in the
earth's gravitational field were observed above a flat neutron optical
mirror by measuring the transmission of ultra-cold neutrons through a narrow
horizontal channel formed by the mirror and an absorber, as a function of
the channel width \cite{Nesvizhevsky/2002}. The nuclear Fermi potential of
the mirror material is given by%
\begin{equation}
U_{\mathrm{F}}=\frac{2\pi \hslash ^{2}}{m_{\mathrm{n}}}Nb,
\end{equation}%
where $m_{\mathrm{n}}$ is the neutron mass, and $b$ is the coherent neutron
scattering length of the mirror nuclei with number density $N$. The energies
of the lowest energy levels of the neutron in the potential well formed by
the earth gravitational potential and the mirror are in the peV range
whereas a typical value for $U_{\mathrm{F}}$ is $100$ neV. It was pointed
out that additional short-range interactions close to the mirror would
modify the transmission pattern. The authors of refs. \cite%
{Nesvizhevsky/2004,Nesvizhevsky/2005} claim that a competitive limit on
non-Newtonian interactions in the nm range follows from the absence of a
bound state due to the hypothetic short-range interaction. We show that this
claim is not valid.

A common parameterisation of short-range modifications of the gravitational
interaction employs a Yukawa-type potential between two masses $m$ and $M$,%
\begin{equation}
V_{G}\left( r\right) =-G\frac{mM}{r}\alpha _{G}\exp \left( -r/\lambda
\right) .  \label{Yukawa}
\end{equation}%
For the extended mass of a flat mirror, integration of eq.(\ref{Yukawa})
provides the effective potential for a neutron situated at distance $z$
outside the mirror with mass density $\rho $,%
\begin{equation}
V_{\mathrm{eff}}\left( z\right) =-U_{0}\exp \left( -z/\lambda \right)
,\qquad U_{0}=2\pi G\alpha _{G}m_{\mathrm{n}}\rho \lambda ^{2}.
\label{V-eff}
\end{equation}%
The authors of refs. \cite{Nesvizhevsky/2004,Nesvizhevsky/2005} replaced the
potential inside the mirror by infinite repulsion and thus consider the
potential well $U=\infty $ for $z\leq 0$, and $U=-U_{0}\exp \left(
-z/\lambda \right) $ for $z>0$, with $U_{0}>0$. The condition for the
absence of a bound state in this well,%
\begin{equation}
U_{0}m_{\mathrm{n}}\lambda ^{2}<0.723\,\hslash ^{2},\qquad \alpha _{G}>0,
\label{condition}
\end{equation}%
follows from transformation of the Schr\"{o}dinger equation for a particle
moving in the exponential pocket to Bessel's differential equation via the
substitution $z=\exp \left( -x/\left( 2\lambda \right) \right) $. From
non-observation of an effect due to a bound state in their experiment \cite%
{Nesvizhevsky/2002}, they conclude from condition (\ref{condition}) an
excluded range of values for the parameter pair $\left( \alpha _{G},\lambda
\right) $. However, their limit is very poor: e.g. for $\lambda =10^{-9}$ m, 
$U_{0}<30$ $\mu \mathrm{eV}$, still allowing for values $U_{0}\gg U_{\mathrm{%
F}}$. For such large values of $U_{0}$ the assumption of an infinite
potential barrier becomes completely unrealistic, since the treatment in
refs. \cite{Nesvizhevsky/2004,Nesvizhevsky/2005} neglects the effect of the
short-range interaction within the mirror. Evaluating the integral leading
to $V_{\mathrm{eff}}\left( z\right) $ also for values $z<0$, we obtain%
\begin{equation}
U\left( z\right) =\left\{ 
\begin{array}{cc}
-U_{0}\left( 2-\exp \left( z/\lambda \right) \right) +U_{\mathrm{F}}\qquad 
& z\leq 0 \\ 
-U_{0}\exp \left( -z/\lambda \right)  & z>0%
\end{array}%
\right. .  \label{potential}
\end{equation}%
Deep inside the mirror ($\left\vert z\right\vert \gg \lambda $) the
short-range interaction thus contributes $-2U_{0}$ to the neutron optical
potential, such that neutrons would no more be totally reflected if $U_{0}$
is larger than half of $U_{\mathrm{F}}$.

Using the potential given in eq.(\ref{potential}), one may exclude values of
the parameter pair $\left( \alpha _{G},\lambda \right) $ which are orders of
magnitude more stringent than those quoted in refs. \cite%
{Nesvizhevsky/2004,Nesvizhevsky/2005}. A sensitive method consists in the
comparison of neutron scattering amplitudes derived from neutron optical and
neutron scattering methods, i.e. for momentum transfer $\hslash q=0$ and $%
\hslash q\neq 0$. A contribution to the scattering amplitude of a nucleus
with mass $M$ from a Yukawa-type potential in Born approximation is given by%
\begin{equation}
f_{G}\left( q\right) =-\frac{m_{\mathrm{n}}}{2\pi \hslash ^{2}}\int
V_{G}\left( r\right) \exp \left( i\mathbf{q}\cdot \mathbf{r}\right) \mathrm{d%
}^{3}r=\frac{2G\alpha _{G}m_{\mathrm{n}}^{2}M}{\hslash ^{2}}\frac{1}{%
q^{2}+\lambda ^{-2}}.
\end{equation}%
Hence, in a measurement at $q=0$, the Yukawa interaction will contribute to
the neutron interaction with a macroscopic body, as given in eq.(\ref%
{potential}) for a flat body with linear extensions $\gg \lambda $, whereas
in a scattering process with $q\gg 1/\lambda $ the effect of $V_{G}\left(
r\right) $ will be strongly suppressed due to the $q$ dependence of $f_{G}$.
This strategy was employed in measurements of the amplitude $f_{\mathrm{ne}%
}\left( q\right) =-ZF\left( q\right) b_{\mathrm{ne}}$ of neutron-electron
scattering, where the known form factor $F\left( q\right) $ of the atomic
electron shell with total charge number $Z$ defines the required $q$ range
for scattering experiments. Koester and coworkers combined measurements of
the total scattering cross section for lead and bismuth at energies $1.26$
eV and $5.19$ eV with neutron optical measurements, from which they derived
a value for the neutron-electron scattering length of $b_{\mathrm{ne}%
}=-\left( 1.32\pm 0.04\right) \times 10^{-3}$ fm \cite{Koester/1986}. An
independent determination of $b_{\mathrm{ne}}$ in measurements of the
neutron scattering asymmetry from noble gases with $q$ values in the range $5
$ nm$^{-1}<q<20$ nm$^{-1}$ by Krohn and Ringo provided a value of $b_{%
\mathrm{ne}}=-\left( 1.30\pm 0.03\right) \times 10^{-3}$ fm \cite{Krohn/1973}%
.

Including the contribution due to the hypothetic gravitational short-range
interaction and defining $b_{G}=-f_{G}\left( 0\right) $, the neutron optical
potential is given by%
\begin{equation}
U=\frac{2\pi \hslash ^{2}}{m_{\mathrm{n}}}N\left( b+Zb_{\mathrm{ne}%
}+b_{G}\right) .  \label{U}
\end{equation}%
The cross section for scattering of neutrons with energy $E_{j}$ of a few eV
is given by (neglecting small corrections) 
\begin{equation}
\sigma \left( E_{j}\right) =4\pi \left( b+Zf_{j}b_{\mathrm{ne}%
}+f_{Gj}\right) ^{2},  \label{sigma}
\end{equation}%
where the number $f_{j}$ describes the influence of the atomic form factor
and $f_{Gj}$ originates from integrating the Yukawa-type gravitational
interaction over the allowed momentum transfers $q$, at incident neutron
energy $E_{j}$. The neutron scattering asymmetry for two values of momentum
transfer $q_{1}$ and $q_{2}$ is given by (not discussing recoil)%
\begin{equation}
A=\frac{\mathrm{d}\sigma }{\mathrm{d}\Omega }\left( q_{1}\right) /\frac{%
\mathrm{d}\sigma }{\mathrm{d}\Omega }\left( q_{2}\right) \simeq 1+2Z\frac{b_{%
\mathrm{ne}}}{b}\left( F\left( q_{1}\right) -F\left( q_{2}\right) \right) +%
\frac{2}{b}\left( f_{G}\left( q_{2}\right) -f_{G}\left( q_{1}\right) \right)
.  \label{ass}
\end{equation}

Allowing a contribution of $b_{G}$ to $U$ ranging up to $Zb_{\mathrm{ne}}$
and focusing on a Yukawa-type interaction with range $\lambda \geq 1$ nm, we
may analyse the situation in leading order in the small amplitudes. For
energies $E_{j}\gtrsim 1$ eV, we have $f_{Gj}\ll Zf_{j}b_{\mathrm{ne}}$
since the values of $\lambda $ considered are much larger than the size of
the atom, and therefore 
\begin{equation}
\sigma \left( E_{j}\right) \simeq 4\pi b\left( b+2Zf_{j}b_{\mathrm{ne}%
}\right) .  \label{sigma-c}
\end{equation}%
Also the last term in eq.(\ref{ass}) becomes negligible compared to the
second one,%
\begin{equation}
A\simeq 1+2Z\frac{b_{\mathrm{ne}}}{b}\left( F\left( q_{1}\right) -F\left(
q_{2}\right) \right) .  \label{ass-c}
\end{equation}%
One thus deals with three observables as functions of the three quantities $%
b $, $b_{\mathrm{ne}}$ and $b_{G}$. From this one may obtain a value,
respectively, a limit for $U_{0}$ for the specified range of $\lambda $.
Indeed such an analysis was already given in 1992 by Leeb and Schmiedmayer 
\cite{Leeb/1992}. The gravitational short-range amplitude appears to leading
order only in $U$, and the contribution of $f_{G}\left( 0\right) $ should
thus appear as a difference between the results for $b_{\mathrm{ne}}$
obtained by measuring the combination of $U$ and $\sigma \left( E_{j}\right) 
$ \cite{Koester/1986}, and via $A$ \cite{Krohn/1973}. From the consistency
of the experimental results the authors of ref. \cite{Leeb/1992} derived $%
b_{G}=\left( -1.6\pm 4.1\right) \times 10^{-3}$ fm, corresponding to $%
U_{0}=\left( -1.4\pm 3.4\right) \times 10^{-11}$ eV. In the present
parametrisation, this leads, within $90$ \% confidence limit, to the
constraint 
\begin{equation}
\left\vert \alpha _{G}\right\vert \lambda ^{2}<530\text{ }\mathrm{m}%
^{2}\quad \text{for }\lambda \geq 1\text{ }\mathrm{nm}.  \label{limit}
\end{equation}%
This limit might become somewhat less stringent if one includes two prior
determinations of $b_{\mathrm{ne}}$ obtained for $^{186}$W and bismuth
published in \cite{Alexandrov/1986}, which are inconsistent with the results
in ref. \cite{Koester/1986}. On the other hand, a more recent precise
determination of $b_{\mathrm{ne}}$ by neutron transmission through liquid $%
^{208}$Pb with the neutron time-of-flight method in the neutron energy range 
$0.08$ to~$800$ eV is in full agreement \cite{Kopecky/1997}. In any case the
limit stated as competitive in ref. \cite{Nesvizhevsky/2004} and as new in
ref. \cite{Nesvizhevsky/2005} is by several orders of magnitude worse: $%
\alpha _{G}\lambda ^{2}<3.4\times 10^{7}$\thinspace $\mathrm{m}^{2}\times
\left( 1\text{\thinspace nm}/\lambda \right) ^{2}$ (note that the comparison
becomes meaningless for $\lambda \gtrsim 10$ nm, where limits more stringent
than (\ref{limit}) were already provided by methods not involving free
neutrons). Note also that the limit (\ref{limit}) is independent of the sign
of $\alpha _{G}$, rendering the separate discussion of a repulsive
short-range interaction in refs. \cite{Nesvizhevsky/2004,Nesvizhevsky/2005}
obsolete.

\end{document}